\input{aipcheck}

\documentclass[
    ,final            % use final for the camera ready runs
  ]
  {aipproc}

\layoutstyle{6x9}

\begin{document}

\title{The Secret XUV Lives of Cepheids: FUV/X-ray observations of Polaris
and $\beta$ Dor}


\author{Scott G. Engle}{
  address={Villanova University, Department of Astronomy \& Astrophysics,
           800 E. Lancaster Ave, Villanova, PA 19085, USA}
  ,altaddress={James Cook University, Centre for Astronomy, Townsville
               QLD 4811, Australia}
}

\author{Edward F. Guinan}{
  address={Villanova University, Department of Astronomy \& Astrophysics,
           800 E. Lancaster Ave, Villanova, PA 19085, USA}
}

\author{Joseph DePasquale}{
  address={Harvard-Smithsonian Center for Astrophysics, 60 Garden St.,
           Cambridge, MA 02138, USA}
}

\author{Nancy Evans}{
  address={Smithsonian Astrophysical Observatory, 60 Garden St., Cambridge,
           MA 02138, USA}
}

\begin{abstract}
We report on the surprising recent discovery of strong FUV emissions in
two bright, nearby Classical Cepheids from analyses of FUSE archival
observations and one of our own approved observations just
prior to the failure of the satellite. Polaris and $\beta$ Dor are
currently the only two Cepheids to have been observed with FUSE, and
$\beta$ Dor is the only one to have multiple spectra. Both Cepheids
show strong C {\sc iii} (977\AA, 1176\AA) and O {\sc vi} (1032\AA,
1038\AA) emissions, indicative of 50,000--500,000 K plasma, well above
the photospheric temperatures of the stars. More remarkably,
$\beta$ Dor displays variability in the FUV emission strengths which
appears to be correlated to its 9.84-d pulsation period. This phenomenon
has never before been observed in Cepheids. The FUV studies are presented
along with our recent Chandra/XMM X-ray observations of Polaris and
$\beta$ Dor, in which X-ray detections were found for both stars. Further
X-ray observations have been proposed to unambiguously determine the
origin and nature of the observed high energy emissions from the targets,
possibly arising from warm winds, shocks, or pulsationally induced
magnetic activity. The initial results of this study are discussed. 
\end{abstract}

\maketitle

%%%%%%%%%%%%%%%%%%%%%%%%%%%%%%%%%%%%%%%%%%%%
%% MAINMATTER
%%%%%%%%%%%%%%%%%%%%%%%%%%%%%%%%%%%%%%%%%%%%

\section{Introduction \& Background}

Classical Cepheids ($\delta$ Cep Variables) are a class of luminous
($\sim$10$^{\rm 3}$--10$^{\rm 4}$ L$_\odot$) yellow (spectral types later
than $\sim$F5) supergiants (luminosity class Ia, Ib \& II) that undergo
regular (periodic, from $<$2 -- $\sim$45 days) radial pulsations. The
lightcurve of a Classical Cepheid (hereafter simply refered to as a Cepheid)
is characterized by a quick rise to maximum brightness followed by a gradual
decline to minimum brightness, giving rise to the ``sawtooth'' shape of its
lightcurve.
However, Cepheids with pulsation periods ranging from $\sim$6.5--20 days
are undergoing the ``Hertzsprung Progression'' -- characterized by a
``bump'' (short increase in brightness) on their lightcurves. 

The ``Secret Lives of Cepheids'' Program is a comprehensive study of Classical
Cepheid behavior \& evolution, pulsation, atmospheres, heating
dynamics and winds that we have been carrying out since 2002. The
program currently consists of $\sim$15 bright Cepheids, covering a wide range
of pulsation properties. We have been obtaining photoelectric photometry
and examining previous observations going back over 100 years (over 2000 years
for Polaris). We are also utilizing IUE/HST and FUSE data to study
Cepheid FUV/NUV characteristics (Engle et al. 2006, 2009), and have recently
obtained Spitzer high resolution
IR spectra of Cepheids to investigate their recently discovered
circumstellar envelopes, possible pulsationally induced winds and/or
magnetic-dynamo chromospheric heating. At X-ray wavelengths we have so far
obtained Chandra/XMM exposures of Polaris, along with recent XMM observations
of $\delta$ Cep and the ``bump
Cepheid'' $\beta$ Dor. More XMM priority-C observations are still awaiting
execution, but the odds of them being carried out are small. We also have
an approved program for Hubble COS UV--FUV
spectrometry of Polaris to search for rapid evolutionary changes. Presently
our program covers almost the entire electromagnetic spectrum, from X-rays
to IR, for this important class of luminous pulsating stars. 

\section{Studying the Upper Atmospheres of Cepheids}

Kraft (1957) made the first comprehensive investigation of Cepheid activity
not completely
confined to the photosphere when he studied the Ca {\sc ii} lines in a large
number (20$+$) of brighter Cepheids. The Ca {\sc ii} {\it HK} lines originate
in plasmas with T $\approx$ 10$^{\rm 4}$ K (similar to the C {\sc iii}
977/1176\AA~lines), similar to that found in the solar
chromosphere. Kraft noted that the Ca {\sc ii} emissions peaked in Cepheids
around $\phi$ $\approx$ 0.9 (just after the Cepheid has begun to expand from
minimum radius -- see Fig. 1a).
Due to the expansion, a shock is expected to pass through the Cepheid
photosphere at this phase, which is sometimes refered to as the ``piston
phase'' in Cepheids.
From this, Kraft concluded that ``the transitory development of Ca {\sc ii}
emission in classical cepheids is associated with the appearance of hot
material low in the atmosphere. These hot gases are invariably linked with
the onset of a new impulse.''

Schmidt \& Parsons (1982; 1984a; 1984b) brought Cepheid activity into the UV
with their rather thorough study of IUE archival spectra. The wavelength
range of IUE covers several important chromospheric (T $\approx$
10$^{\rm 4}$ K) and transition region (T $\approx$ 10$^{\rm 5}$ K)
emission lines. In accord with the results of Kraft (1957), Schmidt \&
Parsons found that the chromospheric emissions were variable, and peaked
just before maximum light (during the {\it piston phase}). Transition
region lines were also found, but were not as strong as chromospheric
emissions and were more easily contaminated by the photospheric continuum in
all but the longest (and best phase-space located) spectra.

Despite the presence of 10$^{\rm 4}$--10$^{\rm 5}$ K emissions (which hint
at the possibility of even hotter plasmas), Cepheids have
long been considered ``X-ray quiet'' stars. B\"ohm-Vitense
\& Parsons (1983) carried out pointed {\it Einstein} observations of three
Cepheids -- $\delta$ Cep, $\beta$ Dor \& $\zeta$ Gem -- but failed to
return a conclusive detection. They reported a {\it possible} detection of 
$\zeta$ Gem, but the counts involved (26 source counts vs. 13 background)
were insufficient to consider $\zeta$ Gem a ``concrete'' X-ray source.

A more recent study by Sasselov \& Lester (1994a; b; c), however,
returned mixed results. Sasselov \& Lester measured the He {\sc i} 10830\AA~
line in a number of Cepheids. This line originates within the chromosphere,
and has a possible (though still debated) connection to coronal emissions.
This line was found to vary over the Cepheids' pulsational phases, in
agreement with earlier studies. Through the phase-lag of the He {\sc i}
curve, though, Sasselov \& Lester were able to calculate the atmospheric 
height at which the line emission originated. For the main target of their
study, $\zeta$ Gem, the He {\sc i} line was found to form at
$\sim$1.3$R_{\rm ceph}$. $\zeta$ Gem has a stellar
radius of $\sim$65$R_\odot$ (Groenewegen 2007), so the chromosphere
of $\zeta$ Gem would extend a further $\sim$20$R_\odot$. With such a large
chromospheric extent, the existence of a corona was put in serious doubt.
However, Sasselov \& Lester still allowed for the existence of coronal
plasmas through atmospheric shocks and acoustic wave dissipation. They also
made theoretical predictions of possible X-ray emissions from Cepheids. For
the $\sim$10-day Cepheid studied ($\zeta$ Gem), they predicted an X-ray
luminosity of $\sim$2.5x10$^{\rm29}$ ergs/sec. 
Sasselov \& Lester also predicted that X-ray emissions could vary by as much
as 2.5x over the pulsational phase, and that Cepheids of longer periods could
have X-ray luminosities as high as $\sim$3x10$^{\rm31}$ ergs/sec. 

\section{FUSE Observations of Classical Cepheids}

The FUV portion of our study was motivated by the results of Schmidt
\& Parsons and by the presence of FUSE archival spectra for Polaris \&
$\beta$ Dor. Schmidt \& Parsons had already detected chromospheric emissions
in all observed Cepheids, and even found limited
transition region emissions in two Cepheids ($\beta$ Dor \& $\zeta$ Gem)
which possessed deeper exposures. They failed to detect transition
region emissions in the shorter period Cepheid $\delta$ Cep, however,
which led them to conclude that perhaps transition regions were
limited to Cepheids of longer periods. Although the IUE data was excellent
in terms of the breadth of its wavelength coverage (including numerous
emission lines) and the number of spectra taken, the specific emissions
of interest are weak, and the photospheric continua of the Cepheids
(F-G supergiants) can easily overwhelm the emission lines. The wavelength
range of the FUSE satellite, although much more narrow than that of IUE,
still includes well known and studied chromospheric and transition region
emission lines, but has the very important added bonus of being free of
any photospheric continuum flux. For these reasons, FUSE was recognized
as a superior satellite for carrying out such a study.

Unfortunately, this recognition came a bit late. We did not know it at
the time, but FUSE was nearing the end of its mission. Our program was
able to carry out one further observation of $\beta$ Dor before FUSE
suffered its fatal malfunction in June, 2007. Thus, the FUSE Cepheid
database remains
two targets strong -- Polaris \& $\beta$ Dor --  and of these targets,
only $\beta$ Dor possesses multiple spectra. However, one of these spectra
was fortunately obtained at a phase where the IUE O {\sc i} emissions
peaked. Fig. 1a shows the FUSE O {\sc vi} \& C {\sc iii}
emissions plotted against the IUE O {\sc i} emissions measured by Schmidt
\& Parsons along with the light, radius and radial velocity curves of
$\beta$ Dor from Taylor \& Booth (1998). As can be seen in the figure, the
chromosphere and transition region of $\beta$ Dor undergo an
excitation at $\phi$ $\approx$ 0.8, after the Cepheid has reached minimum
stellar radius and outward photospheric acceleration has begun. At other
phases, however, weak emissions from these atmospheric layers still remain.
The transition region is apparently less variable than the chromosphere,
indicating that this higher temperature atmospheric layer is either less
affected by the pulsational heating mechanism, or is perhaps more efficient
at storing the pulsational heating. Could this mean that a Cepheid corona is
even less variable, or possible static? Only further data will answer this
question. 

The single FUSE observation of Polaris also contains chromospheric and
transition region emissions, though weaker than $\beta$ Dor. In fact, the
integrated flux of the C {\sc iii} 977\AA~ chromospheric line is $\sim$8x
stronger in $\beta$ Dor at maximum emission than it is in Polaris, but the
O {\sc vi} 1032\AA~ transition region line is only $\sim$3x stronger.
Polaris is a much weaker pulsator than $\beta$ Dor (the light amplitude
of Polaris is only $\sim$0.04-mag when compared to $>$0.6-mag for $\beta$
Dor). Knowing this, the emission line strengths seem to support the
explanation that pulsations have less effect on the heating of the
transition region. However, a single observation of Polaris can not lead
to a conclusive result. Further well-exposed UV/FUV spectra of Cepheids are
needed to understand the true effect of pulsations on their chromospheres
and transition regions.

\section{XMM Observations of Classical Cepheids}

In the spirit of carrying out as comprehensive a study as possible, the
``Secret Lives of Cepheids'' program was extended to cover X-ray wavelengths.
This extension was motivated by the ``$\zeta$ Gem possibility'' and also by 
our discovery of an X-ray detection, in the ROSAT archive, at the position of
the Cepheid Polaris. This detection had, until then, gone completely
unnoticed. The pointed HRI exposure was long enough (just under
8800-seconds) to raise confidence in the existence of an X-ray source at
the coordinates of Polaris (which is, in fact, the position of Polaris Aa
$+$ Polaris Ab - a mid$[$-early$]$ F-type main sequence star $[$dwarf$]$).
Uncertainty in the companion spectral type fostered similar uncertainty in
the true source of the X-rays (Evans et al. 2009). Early F-type dwarf stars
of Pleiades age
(roughly equivalent to the ages of Classical Cepheids) are not known to
produce X-rays (Briggs \& Pye 2003). However, mid F-type dwarfs can produce
X-rays, which meant that the observed X-ray emission could originate in
either the Cepheid Polaris Aa or the companion Polaris Ab (or possibly
from both stars).

To finally prove whether Cepheids produce X-rays, Chandra/XMM time was
successfully proposed for. Observations were carried out for Polaris, 
$\beta$ Dor and $\delta$ Cep -- {\bf all three Cepheids were successfully
detected in X-rays.}
Fig. 1b shows the X-ray energy distributions for the three Cepheids. It is
noteworthy that they display similar distributions, with peak emissions in
the 0.6--0.8 keV range (corresponding temperatures of 7--10 MK). Polaris,
the nearest Cepheid whose detection has the most counts, also displays a
``tail'' of soft X-ray emission not
seen in the other targets. With all three Cepheids having been detected
with similar overall behaviors, we can preliminarily conclude that the
Cepheid Polaris Aa is an X-ray source, and the ``soft tail'' can possibly
be attributed to the companion Polaris Ab. However, other explanations for
this one notable difference in the Cepheid energy distributions are also
very possible. One is that the $\beta$ Dor and $\delta$ Cep exposures
were insufficiently deep to reveal their soft ($<$0.5 keV) emissions, given
the drop in response of the XMM instruments at such energies. Another is
that the Cepheids display different X-ray emissions at different phases
of pulsation. 

The $\beta$ Dor observation is also interesting because of the measured
X-ray luminosity. As stated previously, the X-ray luminosity of $\zeta$ Gem
was theoretically predicted to be $\sim$2.5x10$^{\rm29}$ ergs/sec. $\beta$ Dor
(P $\approx$ 9.84-days) is very similar to $\zeta$ Gem (P $\approx$ 10-days)
and has an observed X-ray lumonosity of 1x10$^{\rm29}$ ergs/sec. This is
close to the predicted value of $\zeta$ Gem, and is based upon one
observation not obtained during the peak chromospheric and transition region
emissions. It will be very interesting to see how the predictions of Sasselov
\& Lester hold up after further data are obtained. 

\begin{figure}
\vspace{-4mm}
  \includegraphics[height=.6\textheight]{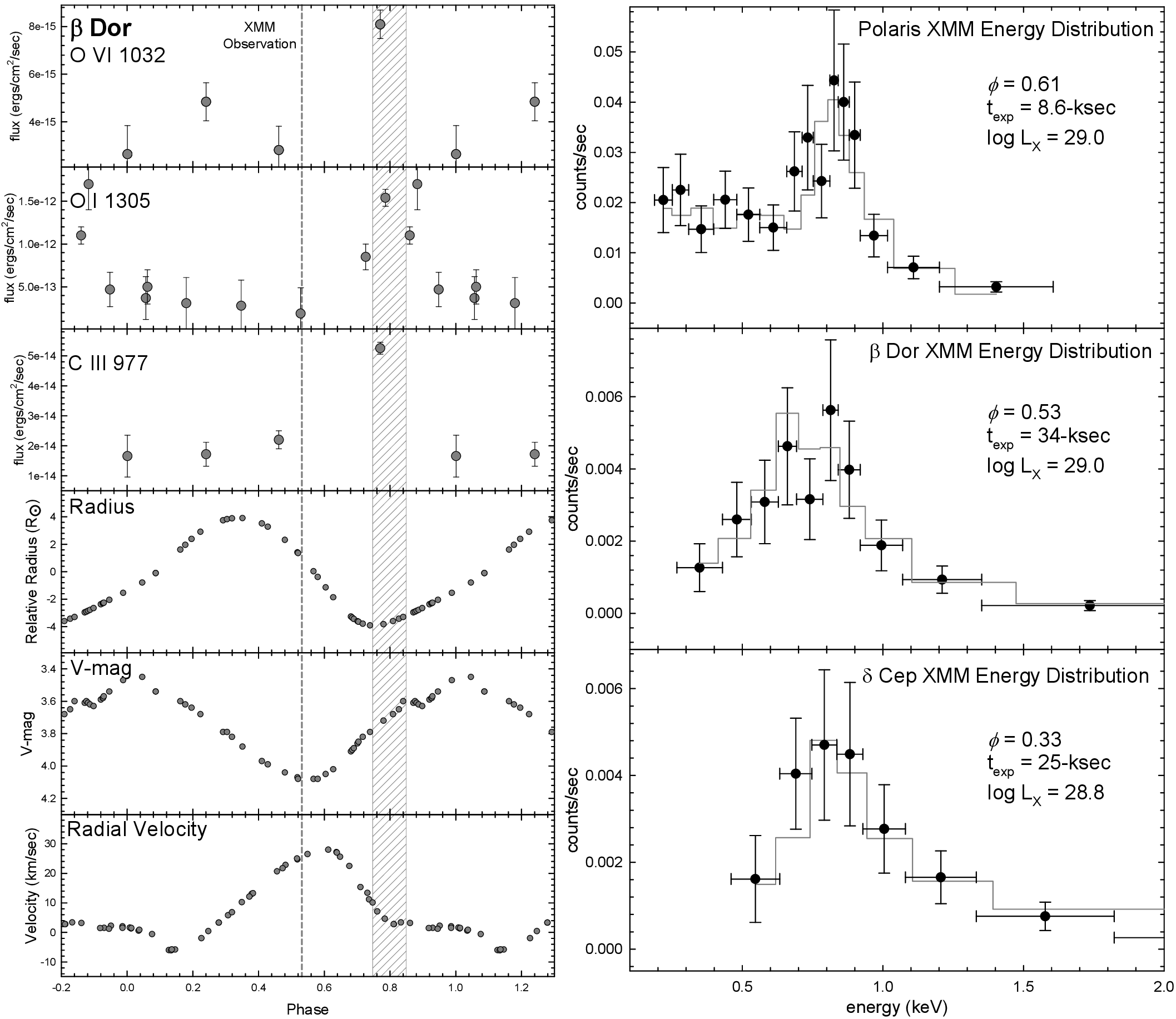}
  \caption{
{{\bf Left (a) --} The FUV (O {\sc vi} and C {\sc iii}) emissions from FUSE and
the O {\sc i} emissions from IUE are plotted for $\beta$ Dor against its 
variations in radius, V-mag and radial velocity over its 9.84-d pulsation
phase. Note the peak in FUV--UV emissions near $\phi \approx$ 0.8 (hashed
region), possibly from pulsationally induced shocks. The dashed line
indicates the phase (0.53) at which our XMM Cycle 7 X-ray observation was
obtained.
{\bf Right (b) --} The X-ray energy distributions (counts/sec vs. energy) are
shown for Polaris, $\beta$ Dor and $\delta$ Cep. It is interesting that the
three models have similar peak emissions in the $\sim$0.6--0.8 keV energy
range. The phases, exposure times and $log$ L$_{\rm x}$ values are given
for each star.}}
\vspace{-4mm}
\end{figure}

\section{Discussion}

In the end, what can the data tell us? Through our use of FUSE archival and
newly obtained spectra (at wavelengths free of photospheric contamination)
we can conclusively say that the upper atmospheres of Cepheids contain
emitting plasmas of T = 10$^{\rm 4}$ -- 10$^{\rm 5}$ K. (We note that,
throughout the paper, we have used ``chromosphere'' and ``transition region''
to describe these emission temperatures. This is meant as a relation of
the emitting temperatures themselves, and not as an implication of the
structure of the Cepheid atmospheres. In all likelihood, the structure of the
Cepheid atmospheres is rather different from that of solar-type stars.)
The emissions are
variable over the pulsational phase of the Cepheid, indicating that they are
linked to the Cepheid pulsations. We are also very excited to report on the
first unambiguous detections of Cepheids at X-ray wavelengths. The three 
Cepheids observed so far -- Polaris, $\beta$ Dor and $\delta$ Cep -- display
similar peak emissions in the $\sim$0.6--0.8 keV energy range, indicating
primary plasma temperatures of 7--10 MK. When taking into account the actual
surface area of the emitting plasmas, the activity of Cepheids is much more
moderate than in solar-type stars (e.g. the X-ray luminosities of Cepheids
are $\sim$100x that of the Sun, but their surface areas are$\sim$2000-4000x).
At this point, insufficient data are
available to make any variability study of the X-ray activity. Such a study
would be necessary to investigate the true extent of pulsational heating in
Cepheid atmospheres, and to also investigate the true similarities and
differences of Cepheid X-ray activity over varying periods of pulsation. The
bottom line is that Cepheids are proving themselves to be more complex than
previously thought and worthy of continued observation. Cepheids are
fundamentally importance to the field of Astronomy, serving as the backbone
of the Extragalactic Distance Scale through their well-studied 
Period-Luminosity Law. Evidently, we still have much to learn in terms of
their true structure and hebaviors. As such, we continue to propose and
hope for further UV--X-ray observations, through which we may finally
understand the true link between stellar pulsations and upper atmospheric
heating processes.

%%%%%%%%%%%%%%%%%%%%%%%%%%%%%%%%%%%%%%%%%%%%
%% Sample figure:
%%
%% The option [height=...] scales the picture to the given height,
%% without it it would be printed at its nominal size
%%%%%%%%%%%%%%%%%%%%%%%%%%%%%%%%%%%%%%%%%%%%

%%%%%%%%%%%%%%%%%%%%%%%%%%%%%%%%%%%%%%%%%%%%
%% SAMPLE TABLE
%%
%% Shows the use of \tablehead and \tablenote
%% macros
%%%%%%%%%%%%%%%%%%%%%%%%%%%%%%%%%%%%%%%%%%%%

%\begin{table}
%\begin{tabular}{lrrrr}
%\hline
%\tablehead{1}{r}{b}{$\beta$ Dor} & & & & \\
%\hline
%\hline
%1982 & 98 & 129 & 620    & 847\\
%1987 & 138 & 176 & 1000  & 1314\\
%1991 & 173 & 248 & 1230  & 1651\\
%1998\tablenote{predicted} & 200 & 300 & 1500  & 2000\\
%\hline
%\tablehead{1}{r}{b}{Polaris} & & & & \\
%\hline
%\hline
%blah & blah & blah & blah & blah \\
%\end{tabular}
%\caption{FUSE Observations of Cepheids}
%\label{tab:a}
%\end{table}

%%%%%%%%%%%%%%%%%%%%%%%%%%%%%%%%%%%%%%%%%%%%%%%%
%% BACKMATTER
%%%%%%%%%%%%%%%%%%%%%%%%%%%%%%%%%%%%%%%%%%%%%%%%

\begin{theacknowledgments}
We gratefully acknowledge support for this project from NASA
grants: Chandra-GO6-7011A, XMM-Newton grant NNX08AX37G, FUSE grant
06-FUSE8-0088 and NSF grant AST05-07542.
\end{theacknowledgments}



\bibliographystyle{aipproc}   % if natbib is available

\bibliography{sample}

\IfFileExists{\jobname.bbl}{}
 {\typeout{}
  \typeout{******************************************}
  \typeout{** Please run "bibtex \jobname" to optain}
  \typeout{** the bibliography and then re-run LaTeX}
  \typeout{** twice to fix the references!}
  \typeout{******************************************}
  \typeout{}
 }

\end{document}